# Effect of "symmetry mismatch" on the domain structure of rhombohedral BiFeO$_3$ thin films

Z. H. Chen,[1,a)] A. R. Damodaran,[1] R. J. Xu,[1] S. K. Lee,[1] L. W. Martin[1,b)]

[1]*Department of Materials Science and Engineering and Materials Research Laboratory, University of Illinois, Urbana-Champaign, Urbana, IL 61801*

Considerable work has focused on the use of epitaxial strain to engineer domain structures in ferroic materials. Here, we revisit the observed reduction of domain variants in rhombohedral BiFeO$_3$ films on rare-earth scandate substrates. Prior work has attributed the reduction of domain variants to anisotropic in-plane strain, but our findings suggest that the monoclinic distortion of the substrate, resulting from oxygen octahedral rotation, is the driving force for variant selection. We study epitaxial BiFeO$_3$/DyScO$_3$ $(110)_O$ heterostructures with and without ultrathin, cubic SrTiO$_3$ buffer layers as a means to isolate the effect of "symmetry mismatch" on the domain formation. Two-variant stripe domains are observed in films grown directly on DyScO$_3$, while four-variant domains are observed in films grown on SrTiO$_3$-buffered DyScO$_3$ when the buffer layer is >2 nm thick. This work provides insights into the role of the substrate – beyond just lattice mismatch – in manipulating and controlling domain structure evolution in materials.

Electronic mail: a) zuhuang@berkeley.edu, b) lwmartin@berkeley.edu

Ferroic materials (*e.g.*, ferroelectrics, ferromagnets, ferroelastics) typically form domains upon cooling from high temperatures in order to minimize the total free energy of the system.[1] Controlling and understanding such domain formation is of critical importance to the advancement of both fundamental studies and applications of these materials as the configuration of domains is critical in determining the ultimate properties of the material.[2,3] In this regard, it is well known that the domain structure of ferroic materials is generally sensitive to the thermal, field, elastic, etc. forces applied to the material. In turn, considerable effort has focused on the use of epitaxial thin-film strain as a way to manipulate, deterministically, the elastic boundary condition and, in turn, gain control of the domain structure of materials.[2-4] Most studies of epitaxial thin-film strain have focused on simple concepts of lattice mismatch (*i.e.*, the sign and magnitude of epitaxial strain),[4] while the effect of "symmetry mismatch" (*i.e.*, arising from a difference in crystal symmetry; particularly, in perovskites where a large variety of octahedral rotations and distortions can produce a range of crystal symmetries derived from a cubic parent structure)[5] between film and substrate has received considerably less attention.

Such concerns become increasingly important when one studies the epitaxy of a film and substrate possessing different symmetries (*e.g.*, cubic and rhombohedral). For instance, studies of rhombohedral ferroic materials, including ferromagnets such as $La_{0.7}Sr_{0.3}MnO_3$[6,7] and the multiferroic $BiFeO_3$,[8-11] have revealed four- and two-variant domain structures for films grown on cubic, $(001)$-oriented and orthorhombic, $(110)_O$-oriented substrates, respectively. (Note that we will use cubic or pseudocubic indices throughout this letter unless otherwise specified and that the subscript "O" denotes orthorhombic indices.) Focusing now on $BiFeO_3$ as a prototypical rhombohedral ferroic material, in the bulk $BiFeO_3$ possesses a rhombohedrally-distorted perovskite structure with a pseudocubic lattice parameter of $a = 3.965$Å, $\alpha = 89.4°$, and space group $R3c$.[12] This, in turn, allows for an $a^-a^-a^-$ antiphase oxygen octahedral rotation (noted in Glazer notation)[13] to occur in this material. In thin-film form, the rhombohedral symmetry is reduced due to the in-plane biaxial strains, resulting in a monoclinic lattice with polarization along $<111>$.[14] Subsequently there are a total of four energetically degenerate structural variants (*i.e.*, elastic domains variants) [Fig. 1(a)], each of which can possess two possible polarization variants,[15] which can give rise to potentially complicated domain pattern with up-to eight possible domain variants.[16] In materials such as BiFeO$_3$ the domain structures not only play a role in the ferroelectric properties,[17] but the domain walls themselves can possess exotic properties, such as electronic conduction.[18] Thus, gaining control over the formation of domain structures is imperative to the ultimate function of the material. This, in turn, has driven numerous studies that have demonstrated deterministic control over the evolution of domain structure[8-10,19] and the hypothesis that anisotropic in-plane lattice parameters of orthorhombic substrates can give rise to selection of a sub-set of the domain variants.[7,9,19] These observations, however, are surprising considering that the strain anisotropy from such orthorhombic substrates is quite small, for instance, the misfit strains between BiFeO$_3$ and DyScO$_3$ $(110)_O$ are -0.3% and -0.4% along the $[001]_O$ and $[1\bar{1}0]_O$, respectively. Such small differences in the magnitude of the anisotropic in-plane strain make it difficult to explain the striking difference in domain variant selection occurring in films grown on cubic and orthorhombic substrates.[10,11] In fact, prior theoretical calculations have suggested that such small anisotropies in the in-plane strain cannot explain a broken degeneracy between the eight polarization variants observed in thin films.[20] Therefore, besides lattice mismatch between the substrate and film, other factors, such as symmetry mismatch, also need to be carefully taken into account to explain the wide-spread observations of domain variant reduction.

In this work, we use the model rhombohedral ferroic material BiFeO3 grown on orthorhombic DyScO3 $(110)_O$ substrates with and without ultrathin, buffer layers of the cubic material SrTiO3 to directly probe the role of symmetry mismatch, rather than just lattice mismatch, plays in determining the domain variant selection in rhombohedral ferroic films. To accomplish this, a series of ultrathin, fully-strained SrTiO3 buffer layers with thicknesses ranging from 0-10 nm were grown on DyScO3 substrates prior to the BiFeO3 growth, thereby producing a series of samples with the same in-plane lattice mismatch, but with varying symmetry mismatch. Subsequently, two-variant stripe domain structures were observed in films grown directly on the DyScO3 substrates, while four-variant domain structures were observed in films grown on SrTiO3-buffered substrates when the buffer layer is > 2 nm thick.

DyScO3 has an orthorhombic structure (space group *Pbnm*, lattice constants $a_0 = 5.440$ Å, $b_0 = 5.717$ Å, and $c_0 = 7.903$ Å)[21] which is the result of cooperative oxygen octahedral rotation in $a^-a^-c^+$ pattern (again in Glazer notation).[13] For a $(110)_O$ substrate, the orthorhombic unit cell can be related to a tilted pseudocubic (monoclinic) unit cell through the following equations: $a = \frac{c_0}{2} = 3.952$ Å, $b = c = \frac{\sqrt{a_0^2+b_0^2}}{2} = 3.947$ Å, $\alpha = 2\tan^{-1}\frac{a_0}{b_0} = 87.2°$, and $\beta = \gamma = 90°$. It is important to note that, the difference between the two in-plane lattice parameters of the substrate is just 0.1%; while the difference between the angles $\alpha$ and $\beta$ ($\gamma$) is over 3%. Unlike DyScO3, bulk SrTiO3 possesses a cubic structure without any octahedral rotations and lattice parameters of $a = b = c = 3.905$ Å at room temperature. Therefore, nominally exact SrTiO$_3$ (001) should not possess any intrinsic anisotropy.

The BiFeO$_3$ films and SrTiO$_3$ buffer layers were grown by pulsed-laser deposition at 700°C in an oxygen pressure of 100 mTorr and 2 mTorr, respectively. The BiFeO$_3$ film thickness was fixed at 50 nm for all samples. To rule out substrate vicinality effects,[22] all the films were grown on nominally exact DyScO$_3$ (110)$_O$ substrates with a miscut angle <0.1°. Detailed structural information was obtained using high-resolution X-ray diffraction (X'Pert MRD Pro, Panalytical) including $\theta - 2\theta$ scans and reciprocal space maps (RSMs). The surface morphology and domain structure were probed using atomic force microscopy (AFM) and piezoelectric force microscopy (PFM) (Asylum Research, Cypher).

A representative AFM topographic image [Fig. 1(b)] and $\theta - 2\theta$ XRD pattern [Fig. 1(c)] of a BiFeO$_3$ film grown on a 10-nm SrTiO$_3$-buffered DyScO$_3$ substrate reveal that the films are single-phase and epitaxial with a smooth surface. The presence of Laue oscillations further confirms that all films have good crystalline quality and smooth surfaces.

Studies of the domain structure reveal uniform out-of-plane PFM contrast suggesting that all domains have uniform out-of-plane polarization orientation.[23] The evolution of the in-plane domain structure with increasing SrTiO$_3$ buffered layer thicknesses is provided [Fig. 2]. In films grown directly on bare substrates [Fig. 2(a)], the domain patterns are dominated by one set of 71° stripe domains with domain walls aligned along [001]$_O$, leading to a BiFeO$_3$ film with only two structural variants. Growth on a ~0.5 nm SrTiO$_3$ buffer layer [Fig. 2(b)], results, again, in primarily two-variant stripe domains. As the thickness of the SrTiO$_3$ buffer layer is increased to ~2 nm, however, a second, orthogonal set of 71° stripe domains with walls along [1$\bar{1}$0]$_O$ appears [Fig. 2(c)]. When the buffer layer thickness increases to 10 nm, the domain patterns are characterized by random combinations of two-variant stripe domains with all four structural variants present is essentially equal fractions [Fig. 2(d)]; which is similar to the domain structure observed in films grown on nominal exact $SrTiO_3$ (001) substrates.[24]

To better understand the domain structure evolution, RSM studies were performed. RSM studies of $BiFeO_3$ films grown on bare $DyScO_3$ substrates about the $DyScO_3$ (pseudocubic $BiFeO_3$) $332_O$- (103-) [Fig. 3(a)], $33\bar{2}_O$- ($\bar{1}03$-) [Fig. 3(b)], $420_O$- (013-) [Fig. 3(c)], and $240_O$- ($0\bar{1}3$-) [Fig. 3(d)] diffraction conditions all reveal that the in-plane lattice parameters of the film are coherently strained to the substrate. Peak splitting occurs in the $h0l$-diffraction condition, but not in the in the $0kl$-diffraction condition. The diffraction studies indicate that only two structural variants ($r_3$ and $r_4$, as defined in Fig. 1(a)) occur in the films grown on bare $DyScO_3$ substrates, consistent with the above PFM results and previous studies.[19] Because of the monoclinic nature of the substrate, the substrate $420_O$-diffraction condition has a different position from the $240_O$-diffraction condition. Likewise, the $BiFeO_3$ 013-diffraction condition also has a different position from the $0\bar{1}3$-diffraction conditoin [Figs. 4(c), (d)], indicating that the $\alpha$ angle of the $BiFeO_3$ pseudocubic unit cell deviates from 90°. A schematic illustration of the two domain motif, as viewed along the $[1\bar{1}0]_O$ [Fig. 4(e)] and $[001]_O$ [Fig. 4(f)], is provided for clarification. Because the substrate has a monoclinic distortion along the $[0\bar{1}1]$ ($[010]_O$), only two structural variants $r_3$ and $r_4$, which have a spontaneous shear distortion along $[\bar{1}\bar{1}1]$ and $[1\bar{1}1]$, respectively, could give rise to a net shear distortion along the monoclinic distortion of the substrate. Therefore, these two variants $r_3$ and $r_4$ are energetically favorable when the rhombohedral films are grown on $(110)_O$ substrates, in order to follow the substrate monoclinic distortion.

RSMs about the DyScO$_3$ (pseudocubic BiFeO$_3$) 332$_O$- ($\bar{1}03$-) and 420$_O$- ($0\bar{1}3$-) diffraction conditions for BiFeO$_3$ films grown on a 10 nm SrTiO$_3$-buffered DyScO$_3$ substrate [Figs. 4(a) and (b), respectively] reveal that both the BiFeO$_3$ and SrTiO$_3$ films are coherently strained to the substrate; that is, both are under anisotropic in-plane strain. Unlike in the films grown directly on DyScO$_3$, however, clear peak splitting occurs in both *h*0*l*- and 0*kl*-diffraction conditions, which typically occurs for BiFeO$_3$ films grown on SrTiO$_3$ substrates.[10,24] This is consistent with the above PFM results that show that all four structural variants are present in the films grown on the SrTiO$_3$-buffered substrate even though the film is still under anisotropic in-plane strain. Therefore, domain variant reduction in the films grown on $(110)_O$ substrates is likely not attributed to the in-plane anisotropic strain alone. Both theoretical and experimental studies have shown that, unlike lattice mismatch, the symmetry mismatch because of different oxygen octahedral tilting systems is normally relieved very rapidly (within only three to four unit cells from the interface).[25-28] In this context, as the thickness of the SrTiO$_3$ buffer layer exceeds 2 nm (or ~5 unit cells) the influence of the oxygen octahedral tilting from the DyScO$_3$ substrate is effectively diminished to a level where it can no longer impact the structure of the SrTiO$_3$ and it takes on a structure devoid of oxygen octahedral tilting and thus there is no monoclinic shear deformation in the topmost SrTiO$_3$ layers. This concept is schematically illustrated [Fig. 4(c) and (d)] and matches what is expected for films grown on cubic SrTiO$_3$ substrates directly where all four structural variants are present in the films when relatively thick cubic buffer layer is inserted.

In conclusion, our experiments directly demonstrate that domain variant reduction in rhombohedral BiFeO$_3$ films grown on $(110)_O$ substrates cannot likely be attributed to anisotropic in-plane strain alone, but appears to be influenced by the presence of a monoclinic distortion, due to oxygen octahedral rotation, in the substrate. Such insights could be applied to other rhombohedral ferroic thin films, such as the ferromagnetic $La_{0.7}Sr_{0.3}MnO_3$ and ferroelectric $PbZr_{1-x}Ti_xO_3$ ($0.06<x<0.47$) to manipulate the domain structures in these materials in a similar fashion. Our results also indicate that it is possible to tune the domain structure of ferroic thin films by engineering octahedral rotation coupling cross interfaces. This is an intriguing observation since increasing the community has been studying how 2-dimensional features, such as surfaces and interfaces, can produce effects that extend well into the bulk of the material. These observations offer yet another example of the power for heteroepitaxial interfaces in directing the evolution of materials.

Z.H.C. and L.W.M. acknowledges the support of the Army Research Office under grant W911NF-10-1-0482, A.R.D. acknowledges the support of the National Science Foundation under grant DMR-1149062, R.X. acknowledges the support of the National Science Foundation under grant DMR-1124696, S.L. acknowledges support of the Air Force Office of Scientific Research under grant MURI FA9550-12-1-0471. Experiments at UIUC were carried out in part in the Materials Research Laboratory Central Facilities.

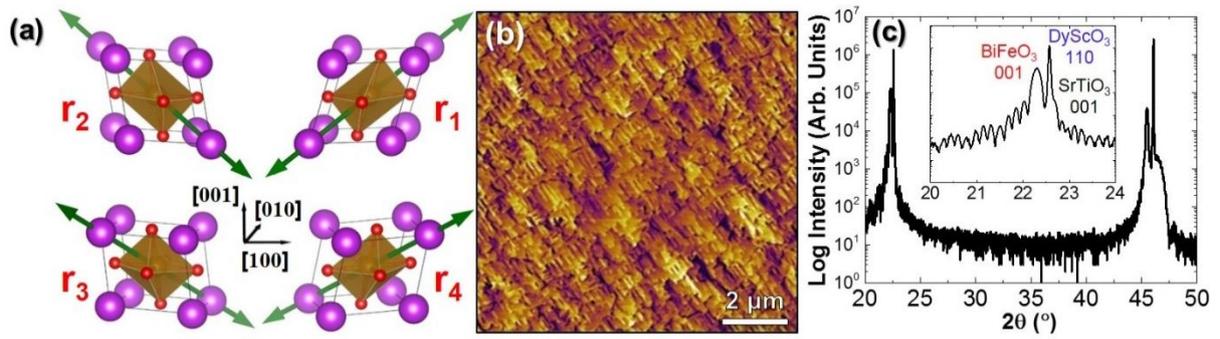

FIG. 1. (a) Illustrations of the all four structural variants in rhombohedral BiFeO$_3$ where the arrows represent the directions of the spontaneous polarization. A typical (b) AFM topographic image and (c) XRD $\theta - 2\theta$ scan of a 50-nm-thick BiFeO$_3$ film grown on 10 nm SrTiO$_3$-buffered DyScO$_3$ $(110)_O$ substrate. Inset shows the enlarged region around 001-diffraction condition of BiFeO$_3$.

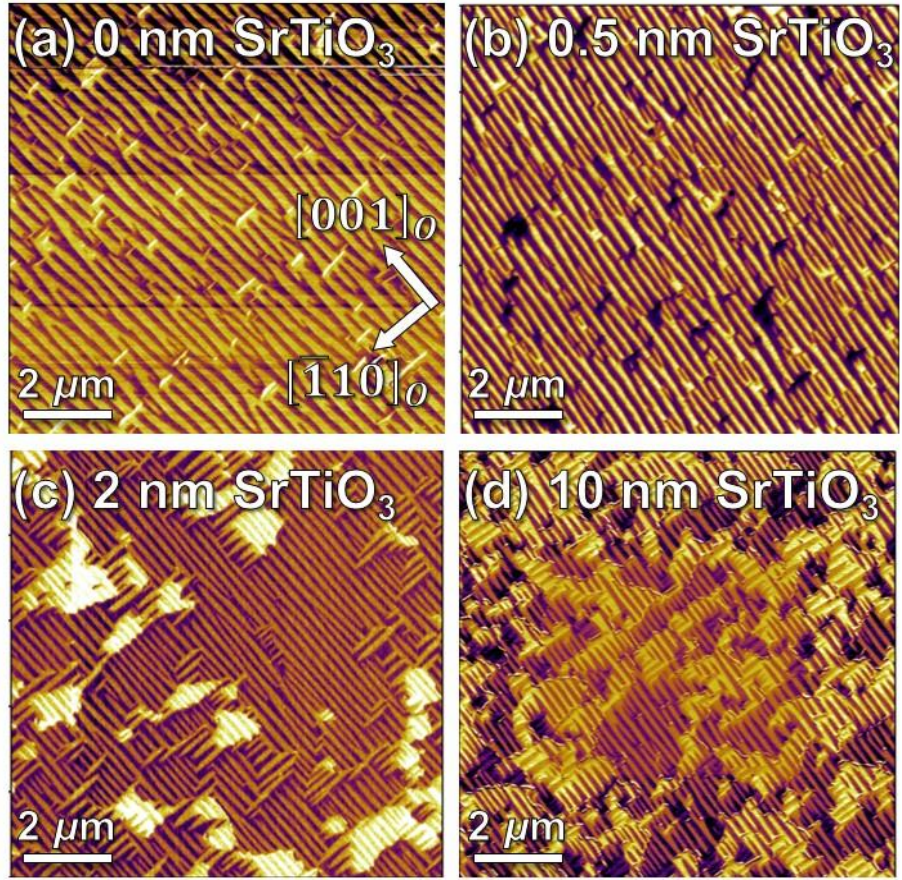

FIG. 2. In-plane PFM images of 50 nm BiFeO$_3$ films grown on DyScO$_3$ (110)$_O$ substrates with SrTiO$_3$ buffer layer thicknesses of (a) 0 nm, (b) 0.5 nm, (c) 2 nm, and (d) 10 nm.

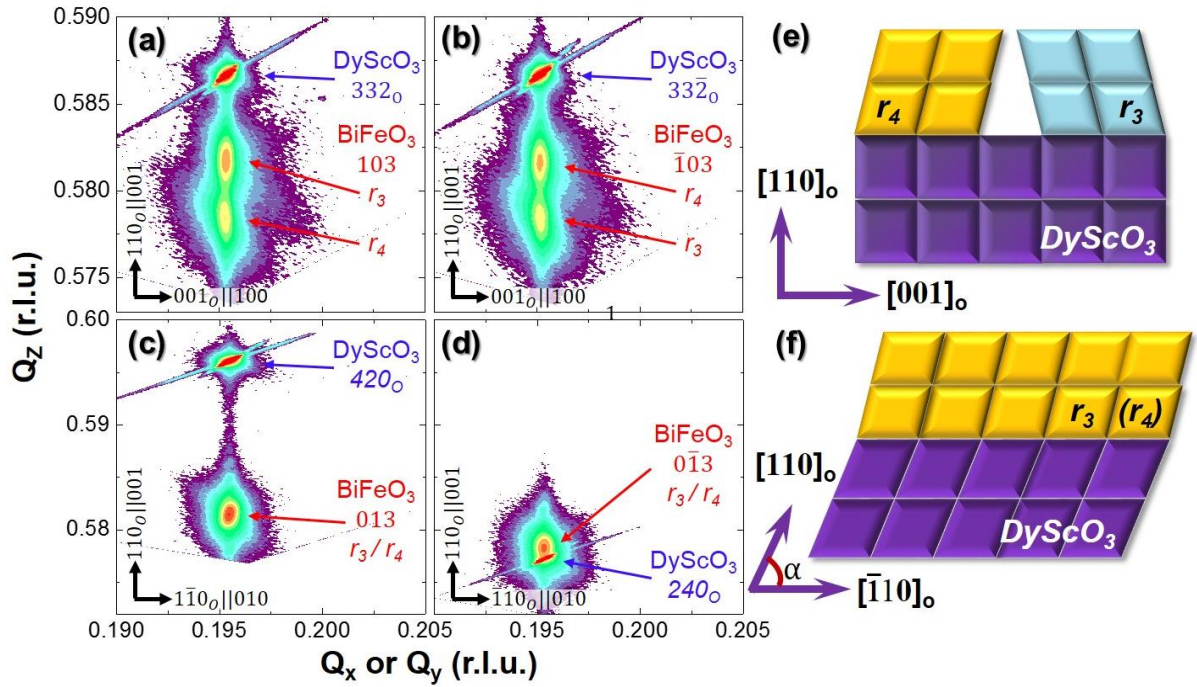

FIG. 3. RSMs of 50 nm BiFeO$_3$ film grown on bare DyScO$_3$ $(110)_O$ substrates about the DyScO$_3$ (pseudocubic BiFeO$_3$) (a) $332_O$- (103-), (b) $33\bar{2}_O$- ($\bar{1}03$-), (c) $420_O$- (013-), and (d) $240_O$- ($0\bar{1}3$-) diffraction conditions. Schematic illustrations of the twinned domain structure along the (e) $[1\bar{1}0]_O$ and (f) $[001]_O$.

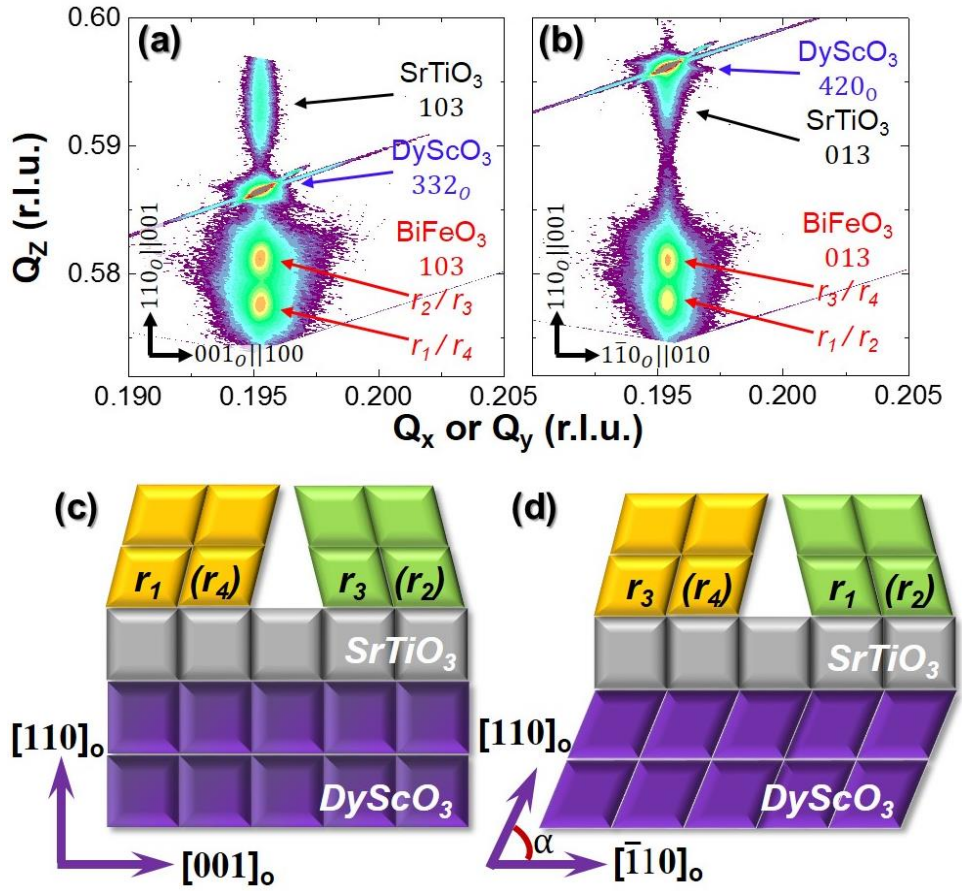

FIG. 4. RSMs of 50 nm BiFeO$_3$ film grown on 10-nm-thick SrTiO$_3$ buffered DyScO$_3$ $(110)_O$ substrate about the DyScO$_3$ (pseudocubic BiFeO$_3$) (a) $332_O$- (103-) and (b) $420_O$- (013-) diffraction conditions. Schematic illustrations of the twinned domain structure along the (c) $[1\bar{1}0]_O$ and (d) $[001]_O$.

# Supplementary Materials for

# Effect of "symmetry mismatch" on the domain structure of rhombohedral BiFeO₃ thin films

Z. H. Chen,[1] A. R. Damodaran,[1] R. J. Xu,[1] S. K. Lee,[1] L. W. Martin[1]

[1]Department of Materials Science and Engineering and Materials Research Laboratory, University of Illinois, Urbana-Champaign, Urbana, IL 61801

1. **AFM topographic image of a 30-nm-thick SrTiO₃ film grown on $(110)_O$ orthorhombic DyScO₃ substrate**

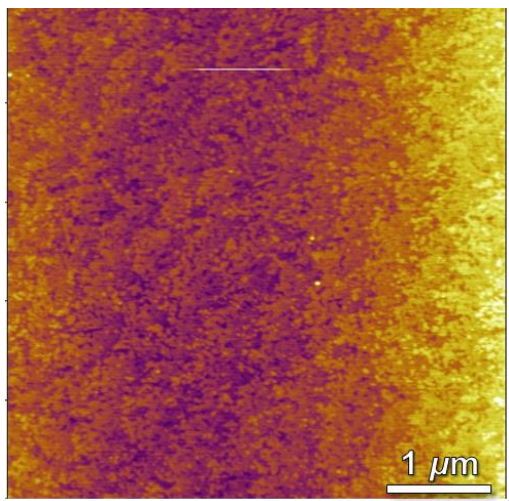

Figure S1. Topographic image of a 30 nm thick SrTiO₃ films grown on the DyScO₃ substrate.

Figure S1 shows a representative topographic image of SrTiO₃ films grown on the DyScO₃ substrate. The film is atomically smooth with root-mean-square (RMS) roughness of ~300 pm. The observed terrace structure reproduces the topography of the substrate.

2. **Reciprocal space mappings (RSMs) of a 60-nm-thick SrTiO₃ film grown on $(110)_O$ orthorhombic DyScO₃ substrate**

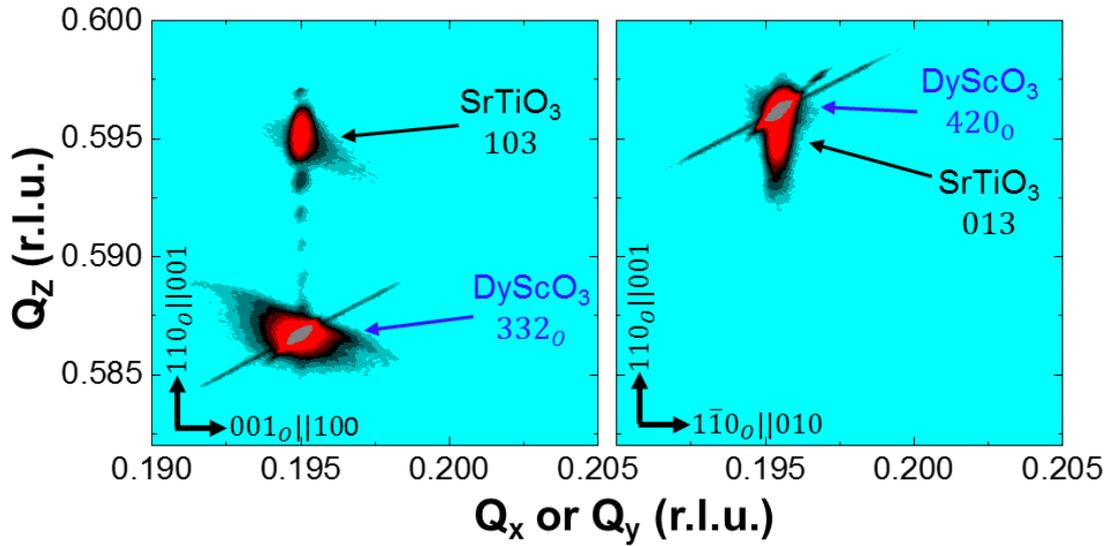

Figure S2. RSMs of a 60 nm SrTiO$_3$ film grown on DyScO$_3$ $(110)_O$ substrate about the DyScO$_3$ (pseudocubic SrTiO$_3$) (a) $332_O$- (103-) and (b) $420_O$- (013-) diffraction conditions.

Figure S2 show RSMs of a 60 nm thick SrTiO$_3$ film grown on the DyScO$_3$ substrate. As can be seen, the reflections from the substrate and film have identical in-plane positions, evidencing that the film is coherently strained along both two in-plane direction. Therefore, we can conclude that all SrTiO$_3$ films with thicknesses below 60 nm should be fully strained to the underlying substrate.

## 3. Out of plane PFM image of the BiFeO$_3$ film grown on $(110)_O$ orthorhombic DyScO$_3$ substrate

Figure S3 show a typical out-of-plane PFM image of the BiFeO$_3$ films. No clear contrast was observed, indicating that all domains have uniform out-of-plane polarization orientation.

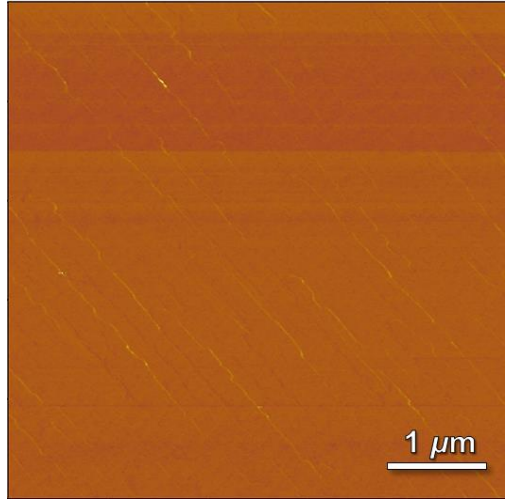

Figure S3. Out-of-plane PFM image of a 50 nm thick BiFeO$_3$ film grown on bare DyScO$_3$ $(110)_O$ substrate.